\documentclass[preprint,aps,nofootinbib]{revtex4}

\usepackage{graphicx}% Include figure files
\usepackage{dcolumn}% Align table columns on decimal point
\usepackage{bm}% bold math
\usepackage{epsfig}

\usepackage{amsmath}
\usepackage{amssymb}
\usepackage{color}
\usepackage{hyperref}

\def\fun#1#2{\lower3.6pt\vbox{\baselineskip0pt\lineskip.9pt
  \ialign{$\mathsurround=0pt#1\hfil##\hfil$\crcr#2\crcr\sim\crcr}}}

\def\simgt{\stackrel{>}{{}_\sim}}

\input epsf

\newcommand{\be}{\begin{equation}}
\newcommand{\ee}{\end{equation}}
\newcommand{\bea}{\begin{eqnarray}}
\newcommand{\eea}{\end{eqnarray}}

\begin{document}

\begin{flushright}
ANL-HEP-PR-12-35
\end{flushright}

\vspace*{1cm}

\title{Implications of a Modified Higgs to Diphoton Decay Width}

\vspace*{0.2cm}

\author{
\vspace{0.5cm} 
Marcela Carena$^{a,d,e}$, Ian Low$^{b,c}$,  and Carlos E.~M.~Wagner$^{b,d,e}$ }
\affiliation{
\vspace*{.2cm}
$^a$  \mbox{Fermi National Accelerator Laboratory, P.O. Box 500, Batavia, IL 60510}\\
$^b$ \mbox{High Energy Physics Division, Argonne National Laboratory, Argonne, IL 60439}\\
$^c$ \mbox{Department of Physics and Astronomy, Northwestern University, Evanston, IL 60208} \\
$^d$  \mbox{Enrico Fermi Institute, University of Chicago, Chicago, IL 60637}\\
$^e$  \mbox{Kavli Institute for Cosmological Physics, University of Chicago, Chicago, IL 60637}
\vspace*{0.8cm}}

\begin{abstract}
\vspace*{0.5cm}
Motivated by recent results from Higgs searches at the Large Hadron Collider, we consider possibilities to enhance the diphoton decay width of the Higgs boson over the Standard Model expectation, without modifying either its production rate or the partial widths in the $WW$ and $ZZ$ channels. Studying effects of new charged scalars, fermions and vector bosons, we find that significant variations in the diphoton width may be possible if the new particles have  light masses of the order of  a few hundred GeV and sizeable couplings to the Higgs boson. Such couplings could arise naturally if there is large mass mixing between two charged particles that is induced by the Higgs vacuum expectation value. In addition,  there is generically also a shift in the $Z\gamma$  partial width,  which in the case of new vector bosons tends to be of similar magnitude as the shift in the diphoton partial width, but  smaller in other cases. Therefore simultaneous measurements in these two channels could reveal properties of  new  charged particles at the electroweak scale.
\end{abstract}

\maketitle

\section{Introduction}

The Standard Model (SM) provides an excellent description of all observed phenomena at high energy physics experiments.   The gauge structure of the SM forbids the presence of explicit masses for the fundamental fermions and gauge bosons.  These masses are therefore associated with the breakdown of the electroweak symmetry.   The spontaneous breaking of the gauge symmetry in the SM is engineered via the introduction of a fundamental scalar, transforming in the fundamental representation of the $SU(2)_L$ group, and leads to the presence of a new physical degree of freedom, the Higgs boson, with no electromagnetic or color charges, and with tree-level couplings to the fundamental fermions (massive gauge bosons) which are proportional to (the square of) their masses. 

Searches for a SM-like Higgs boson are underway at the Tevatron and Large Hadron Collider (LHC) experiments.  The Tevatron experiments search  for a Higgs produced in the dominant gluon fusion channel and decaying into the weak gauge bosons, as well as the associated production of a Higgs with weak gauge bosons and decaying into bottom quarks. Due to the limited Tevatron energy, it is sensitive to Higgs boson masses smaller than about 200~GeV. No excess of events were observed in the high mass range, but a broad excess consistent with a SM-like Higgs decaying into bottom quarks was observed for masses in the 115-135 GeV range \cite{TEVNPH:2012ab}.  The statistical significance of the observed excess is, however,  less than 3~$\sigma$ and therefore not sufficient to claim evidence for the Higgs boson.  The Tevatron run is over and an increase in significance of the Higgs signal may only come from  further refinement in the search efficiencies, for the data already analyzed.

The LHC experiments, on the other hand, have sensitivity for Higgs bosons produced in gluon fusion and decaying into weak gauge bosons, for Higgs masses from about 120 GeV up to a mass close to 600~GeV. No significant excess has been seen for masses above 129~GeV and both LHC experiments therefore exclude the presence of a SM Higgs boson in the 129--539~GeV range at the 95\% confidence level \cite{higgscombine}. The LHC is also sensitive to the decay of Higgs bosons into diphoton states, for masses in the 114--130~GeV range. Quite intriguingly, both experiments observed an excess of events in this channel, consistent with the production of a Higgs boson with a mass of about 125~GeV, with a local significance which is close to 3$\sigma$ \cite{ATLAS:2012ad, Chatrchyan:2012tw}. There is also an excess in the production of pairs of $Z$ gauge bosons at the ATLAS experiment in this mass range \cite{ATLAS:2012ac}. A similar search at the CMS experiment reveals a somewhat less significant result \cite{Chatrchyan:2012dg}.   These two search channels provide the best Higgs mass resolution and therefore are powerful in probing the presence of a Higgs boson in the narrow mass range around 125 GeV. A naive combination of the results of both experiments seems to reveal a central value of $ZZ$ production with a rate similar to the SM one, while the central value of the diphoton production rate appears to be  enhanced by 1.5 to 2 times the SM one. The excesses seen in the $h \rightarrow \gamma \gamma$ and $h \rightarrow ZZ \rightarrow 4\ell$ channels are somewhat offset by the more background-like outcome in the $h \rightarrow WW$ searches \cite{ATLASWW}.
More statistics would be needed to determine if these results are significant or are just the product of a statistical fluctuation.

Motivated by these results, we shall investigate the possibility that the diphoton rate is enhanced, and that this enhancement is entirely due to an increase of the partial diphoton decay width of the Higgs, but without significantly varying the total width or production cross sections with respect to their SM values.  Since the Higgs coupling to photons is induced at the loop-level,  such an enhancement of the diphoton decay width demands the presence of colorless charged particles with significant couplings to the Higgs boson that will add to the dominant SM contribution from the $W^\pm$  boson loop.  On the other hand, SM fermions which receive their mass via a Yukawa coupling to the Higgs,  give subleading corrections  which suppress the diphoton partial width. Therefore, a modified diphoton rate suggests the presence of new charged particles and we will see  that an enhanced width in this  channel  points to an interesting structure of the couplings of the Higgs boson to these new charged particles.

A large number of works have studied effects of new particles in the diphoton decay widths of the Higgs as well as in the gluon fusion production channel  \cite{Petriello:2002uu, Djouadi:1998az, Low:2009di, Low:2009nj, Carena:2011aa,Carena:2012gp}, which is also a loop-induced process.\footnote{In this article we shall not analyze the effects of Higgs mixing, as the ones present in models with extended Higgs sectors~(see, for example, Ref.~\cite{extended}).} However, here we wish to emphasize the generic properties of  light charged particles leading to an enhancement of the Higgs diphoton width, as well as their physical implications, beyond the shift in the diphoton rate. For example, the LEP experiments put a strong constraint on the presence of charged particles with mass lower than about 100 GeV \cite{lepstau,lepschargino}, and avoiding these bounds while keeping a significantly increased diphoton rate may imply an enhanced coupling to the Higgs boson.   More interestingly, we point out that any change in the diphoton width is  accompanied by a corresponding modification in the $Z\gamma$ channel, since any charged particle has a non-vanishing coupling to the $Z$ boson generically, and that different new particles give rise to different correlation patterns between these two channels. These particles may induce corrections to the precision electroweak observables and yield new minima in the Higgs potential at tree-level or via radiative corrections. However, these problems can be remedied in a complete model, and given that more data will be available in the near future, we would like to work in a model-independent fashion and shall not be concerned with these indirect constraints.  Instead, we argue that indirect evidence  for new light particles in the $\gamma\gamma$ and $Z\gamma$ \cite{Gainer:2011aa} channels would point to a rich structure of new particles at the TeV scale.

This article is organized as follows : in Section \ref{sect:II} we develop a general understanding of the deviations in the Higgs coupling to photons due to presence of new charged particles.  In Section \ref{sect:III} we discuss specific examples associated with particles of spin zero, spin one-half, and spin one, while in Section \ref{sect:IV} we work out the correlations between $\gamma\gamma$ and $Z\gamma$ partial widths. Then we conclude in Section \ref{sect:V}. In the Appendix we collect expressions for the loop functions used in the calculations.

\section{General features of the HIggs to diphoton decay width}
\label{sect:II}

%%%%%%%%%%%%%%%%%%%%%%%%%%%%%
\begin{figure}[t]
\includegraphics[scale=0.7, angle=0]{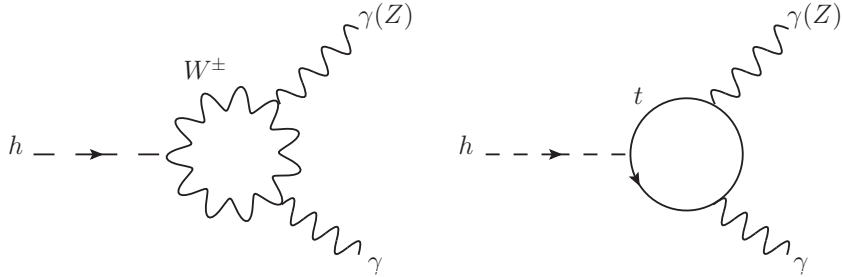}  
\caption{\label{fig0}{\em SM contributions to Higgs decays in the $\gamma\gamma$ and $Z\gamma$ channel.
}}
\end{figure}
%%%%%%%%%%%%%%%%%%%%%%%%%%%%%%%%  

 In the SM the leading contribution to the Higgs coupling to diphoton is the $W^\pm$ boson loop, which is at least four times larger than the next-to-leading contribution from the top quark loop. The Feynman diagrams are shown in Fig.~\ref{fig0}, where the same diagrams also constitute the dominant contributions to  the Higgs coupling to $Z\gamma$. The analytic expression for the diphoton partial width reads \cite{Ellis:1975ap, Shifman:1979eb}
\be
\label{eq:exact}
\Gamma(h\to \gamma\gamma)=\frac{G_F \alpha^2 m_h^3}{128\sqrt{2}\pi^3}\left|A_1(\tau_W)+ N_c Q_t^2  A_{1/2}(\tau_t) \right |^2 \ ,
\ee
where $G_F$ is the Fermi constant, $N_c=3$ is the number of color, $Q_t=+2/3$ is the top quark electric charge in units of $|e|$, and $\tau_i\equiv 4m_i^2/m_h^2$, $i=t, W$. Below the $WW$ threshold, the loop functions  for spin-1 ($W$ boson) and spin-1/2 (top quark)  particles are given by Eqs.~(\ref{eq:loop1}) and (\ref{eq:loop2}) in the Appendix.

In the limit that the particle running in the loop has a mass much heavier than the Higgs, we have
\be
\label{eq:limit}
A_1 \rightarrow -7  \ , \qquad  N_c Q_t^2\, A_{1/2} \rightarrow \frac{4}3 N_c Q_t^2 \ .
\ee
For a Higgs mass below the $WW$ threshold, the $W$ boson contribution is always dominant and monotonically decreasing from $A_1=-7$ for very small Higgs masses to $A_1\approx -12.4$ at the threshold, while the top quark contribution is well-approximated by the asymptotic value of $(4/3)^2\approx 1.78$. If we consider a Higgs mass at 125 GeV, the $W$ and top contributions are
\be
m_h=125 \ \ {\rm GeV}: \ A_1=-8.32 \ , \quad  N_c Q_t^2 A_{1/2}=1.84\ .
\ee
We will investigate effects on the diphoton width from adding new colorless charged particles of spin-0, spin-1/2, and spin-1, which would interfere with the SM contributions. In particular, we are interested in investigating under which circumstances  the di-photon partial width could be significantly enhanced . 

We begin by re-writing the diphoton decay width in terms of the Higgs coupling to the particles in the loop:
\be
\label{eq:exact2}
\Gamma(h\to \gamma\gamma)=\frac{ \alpha^2 m_h^3}{1024\pi^3}\left|\frac{g_{hVV}}{m_V^2} Q_V^2 A_1(\tau_V)+ \frac{2g_{hf\bar{f}}}{m_f} N_{c,f} Q_f^2  A_{1/2}(\tau_f) +  N_{c,S} Q_S^2 \frac{g_{hSS}}{m_S^2} A_0(\tau_S) \right |^2 \ ,
\ee
In the above the notation $V$, $f$, and $S$ refer to generic spin-1, spin-1/2, and spin-0 particles, respectively.  $Q_V$, $Q_S$  and $Q_f$ are the electric charges of the vectors, scalars and fermions in units of $|e|$, $N_{c,f}$ and $N_{c,S}$ are the number of fermion and scalar colors and the scalar loop function $A_0$ is defined in Eq.~(\ref{eq:loop3}) in the Appendix. $A_0$ approaches 1/3 for infinitely heavy scalar masses in the loop.
For the standard model  case, the $W$ boson and top quark couplings to the Higgs are given by  $g_{hWW}=g^2 v/2$ and $g_{ht\bar{t}}=\lambda_t/\sqrt{2}$, and
\be
\label{eq:smcoupling}
\frac{g_{hWW}}{m_W^2}=\frac{2g_{ht\bar{t}}}{m_t}  = \frac{2}v \ ,
\ee
where $v\approx 246$ GeV is the Higgs vacuum expectation value (VEV). Using Eq.~(\ref{eq:exact2}) one could easily include new loop contributions in the diphoton decay width.

To understand the pattern of deviations in the diphoton width, it is instructive to use the low-energy Higgs theorems \cite{Ellis:1975ap,Shifman:1979eb} to derive leading contributions to the diphoton decay width from new heavy particles, although in the specific examples considered later we always include the finite mass effect. The theorems relate the partial decay width to the $\gamma\gamma$ two point functions. As a result, the leading contribution in the $h\to \gamma\gamma$ partial width can be obtained from the knowledge of  one-loop QED beta functions. More specifically, in the presence of charged heavy particles, the QED effective Lagrangian at one-loop order is given by
\be
\label{eq:Lgaga}
{\cal L}_{\gamma\gamma} = -\frac14 F_{\mu\nu} F^{\mu\nu} \sum_i \frac{b_i e^2}{16\pi^2} \log \frac{\Lambda^2}{m_i^2} + \cdots \ ,
\ee
where $m_i$ is the mass of the $i$th particle, $\Lambda$ is an ultraviolet cutoff, and the beta function coefficients $b_i$ are \cite{Ellis:1975ap,Shifman:1979eb}
\bea
\label{eq:beta-functions}
b_{1/2}&=& \frac43 N_{c,f} Q_f^2  \quad \text{for  a Dirac fermion} \ , \\
\label{eq:beta-functions1}
b_1&=& -7 \quad\quad\quad \ \  \text{for the $W$ boson}\ , \\
\label{eq:scalarbeta}
b_0&=& \frac13 N_{c,S} Q_S^2  \quad \ \text{for a charged scalar}\ .
\eea 
The $-7$ coefficient for the $W$ boson can be understood as the sum of $-22/3$, which is the beta function coefficient for non-abelian gauge bosons, and $+1/3$, which comes from the scalar (longitudinal) components of the massive gauge bosons \cite{Ellis:1975ap,Shifman:1979eb}.

Since we are interested in an enhanced $\gamma\gamma$ width without changing the Higgs production rate, we only consider new particles carrying no color charges and set $N_c=1$ henceforth.\footnote{In the gluon fusion production of the Higgs, if the amplitude from a new colored particle is approximately twice as large as that from the SM top but with an opposite sign, the resulting amplitude simply changes sign and the production cross section could remain roughly  the same. This way one could enhance the diphoton decay width without changing the production rate using a new colored particle. This scenario has the same effect as flipping the sign of the linear $h$-$t$-$t$ coupling, relative to the top mass, using higher dimensional operators and is clearly very special. We do not consider this possibility further in this work.}  Moreover, if the mass of the new particle depends on the Higgs VEV,\footnote{The new particle does not have to receive all of its mass from the Higgs VEV, but only some of it is sufficient.} $m_i \to m_i(h)$, and is much heavier than $m_h$, we can integrate out the heavy new particle and describe the Higgs coupling to two photons using an effective Lagrangian in a $1/m_i$ expansion. In the end the $h\gamma\gamma$ coupling is readily obtained by making the substitution $h\to h+v$ in Eq.~(\ref{eq:Lgaga}) and expand to linear order in $h$:
\be
\label{eq:Lhgaga1}
{\cal L}_{h\gamma\gamma} = \frac{\alpha}{16\pi}\frac{h}{v} \left[ \sum_i  2 b_i \frac{\partial}{\partial \log v} \log m_i(v)\right] F_{\mu\nu} F^{\mu\nu}  \ .
\ee
In terms of the notation in Eq.~(\ref{eq:exact2}), 
\be
\label{eq:generalcoup}
\frac{g_{hVV}}{m_V^2}= \frac{\partial}{\partial v} \log m_V^2(v)\ , \quad  \frac{2g_{hf\bar{f}}}{m_f}=\frac{\partial}{\partial v} \log m_f^2(v)\ , \quad \frac{g_{hSS}}{m_S^2}= \frac{\partial}{\partial v} \log m_S^2(v)\ .
\ee
In the limit of heavy masses, the exact result in Eq.~(\ref{eq:exact2}) is in full agreement with Eq.~(\ref{eq:Lhgaga1}).

When there are multiple particles carrying the same electric charge, one can write down a slightly more general expression
\be
\label{eq:Lhgaga2}
{\cal L}_{h\gamma\gamma} = \frac{\alpha}{16\pi}\frac{h}{v} \left[ \sum_i   b_i \frac{\partial}{\partial \log v} \log\left(\det {\cal M}_{F,i}^\dagger {\cal M}_{F,i}\right)
+  \sum_i   b_i \frac{\partial}{\partial \log v} \log\left(\det {\cal M}_{B,i}^2\right)\right]
F_{\mu\nu} F^{\mu\nu}  \ ,
\ee
where ${\cal M}_{F,i}$ and ${\cal M}_{B,i}$ are the mass matrices of all particles carrying the same electric charge and spin, and $F$ and $B$ denote fermions and bosons. This expression allows for the possibility that there could be mass mixing between particles. In particular, we will be focusing on scenarios where the mass mixing is induced  after the electroweak symmetry breaking, which occurs in many theories beyond the SM.

The form of the effective Higgs coupling to two photons in Eq.~(\ref{eq:Lhgaga2}) makes it straightforward to understand the pattern of deviation from SM expectations in the presence of extra particles running in the loop. As a simple example, we consider the addition of two new fermions. The same consideration applies to scalars by simple substitutions of mass matrices.  In this case,  the mass matrix is a $2\times 2$ matrix,
\be
{\cal M}_f^\dagger {\cal M}_f = \left( \begin{array}{cc}
           {m}_{11}^2 &  m_{12}^2 \\
           m_{12}^{*\, 2} & {m}_{22}^2 
           \end{array} \right)\ ,
\ee
from which the $h\gamma\gamma$ coupling is determined from Eq.~(\ref{eq:Lhgaga2}) by
\bea
\label{eq:generalmix}
&& \frac{ \alpha \: b_{1/2}}{16 \pi} \frac{\partial}{\partial v} \log\left(\det {\cal M}_f^\dagger {\cal M}_f \right) \nonumber \\
&& = \frac{ \alpha \; b_{1/2}} {16 \pi \left(m_{11}^2 m_{22}^2 - \left|m_{12}^2\right|^2\right)}
 \left( m_{11}^2 \frac{\partial}{\partial v}  m_{22}^2 +m_{22}^2 \frac{\partial}{\partial v}  m_{11}^2 -   \frac{\partial}{\partial v}\left|m_{12}^2\right|^2 \right)\ .
\eea
A few comments are in order. First we assume no mass mixing, $m_{12}^2=0$. In this case it is interesting to consider the situation where both particles receive all of their masses from electroweak symmetry breaking, $m_{ii}^2 = d_i v^2$, where $d_i>0$ as required by the condition of positivity  of the mass. Then the first two terms in Eq.~(\ref{eq:generalmix}) contribute with the same sign. This argument suggests that adding a fourth generation quark and/or lepton would always amplify the effects of SM quarks and/or leptons in the loop-induced decay of the Higgs, which implies a  reduction in the diphoton decay width.\footnote{One can apply the same argument to gluon fusion production of the Higgs and arrive at the well-known result that a fourth generation quark will amplify the effect of the SM quarks, thereby enhancing the production cross section with respect to the SM.}  When turning on the mixing parameter $m_{12}^2$, there are two possibilities. The first is that the off-diagonal mixing $m_{12}^2$ is independent of the Higgs VEV, as may be the case when the two particles have the same $SU(2)_L\times U(1)_Y$ quantum numbers, then $m_{12}^2$ only enters in the denominator in Eq.~(\ref{eq:generalmix}), which  must be positive-definite in order to avoid massless and/or tachyonic states. Thus the pattern of interference effect in this case is independent of the off-diagonal entry $m_{12}^2$. The other possibility may occur when the two particles belong to different representations of $SU(2)_L$, such as an $SU(2)_L$ doublet and an $SU(2)_L$ singlet, respectively. Then, $m_{12}^2 \propto v$ and the interference pattern is quite sensitive to the off-diagonal mixing due to the minus sign in front of it in the numerator in Eq.~(\ref{eq:generalmix}). An enhancement of the diphoton coupling to the Higgs may be obtained in this case, since the contribution of the off-diagonal term has the same sign as the leading SM contribution from the $W$ loop. Moreover, the deviation with respect to the SM rate could be significant when the mixing is large, which is well-known in the context of squark and slepton contributions to Higgs production and decays in supersymmetry \cite{Low:2009nj, Carena:2011aa, Djouadi:1998az}.

Eq.~(\ref{eq:Lhgaga2}) also suggest a possible connection between the interference pattern in the diphoton width and the cancellation of one-loop Higgs quadratic divergence, which was studied in Ref.~\cite{Low:2009di} in the context of gluon fusion production of the Higgs boson. The one-loop quadratic divergence in the Higgs mass is contained in the Coleman-Weinberg potential \cite{Coleman:1973jx},
\be
\label{eq:supertrace}
\frac1{16\pi^2}\, \Lambda^2 \ {\rm Str}\  {\cal M}^\dagger {\cal M}   \ ,
\ee
and can be obtained from the $H^\dagger H$ term in the mass matrix squared, after turning on the Higgs as a background field. We have used the super-trace notation in Eq.~\ref{eq:supertrace}) to incorporate scenarios where particles with different spins could contribution to the quadratic divergences. Assuming no mass mixing between particles, we see that a scalar worsening the SM top quadratic divergence and a fermion canceling the SM top quadratic divergence will both interfere destructively with the SM top quark contribution in the diphoton amplitude. The interference with the leading contribution, which comes from the $W$ boson loop, is thus constructive and tends to enhance the diphoton width. The reason for the different pattern between scalar and fermion is due the fact that they have opposite sign in the super-trace in Eq.~(\ref{eq:supertrace}) while in the QED one-loop beta functions they have the same sign. From this argument it is also easy to see that a four-generation lepton has the tendency to reduce the diphoton decay width, since it only worsens the SM top quadratic divergence in the Higgs mass.

\section{Specific examples}
\label{sect:III}

Next we consider specific examples where the $h\to \gamma\gamma$ partial width can be enhanced significantly over the SM expectations. 

\subsection{A new $W'$ boson}

\begin{figure}[t]
\includegraphics[scale=0.5, angle=0]{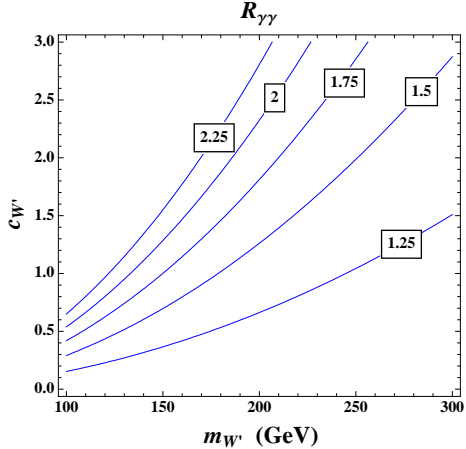}  
\caption{\label{fig1}{\em  Contours of constant diphoton partial width, normalized to the SM value, shown  as a function of $c_{W'}$, the $W'$ coupling strength to the Higgs as defined in Eq.~(\ref{mW'}), and the new $W'$ boson mass.
}}
 \end{figure}

Given that the SM contribution is dominated by the $W$ boson loop, one could add a  $W'$ boson, defined as the $T^3=\pm 1$ component of an $SU(2)_L$ triplet, which has the following mass when turning on the Higgs VEV,
\be
\label{mW'}
m_{W'}(v)^2 = m_{W0}^2 +  c_{W'} \, m_W^2 \ , \qquad c_{W'} > 0 \ ,
\ee 
where $m_{W}^2=g^2v^2/4$ is the mass of the $W$ boson in the SM and we assume $m_{W0}^2$ is independent of $v$. The coefficient $c_{W'}$ parametrizes the coupling of the $W'$ boson with the Higgs, which we take as a free parameter. The only requirement is $c_{W'} > 0$ so that the $W'$ boson loop interferes constructively with the $W$ boson loop, leading to an enhanced diphoton partial width. In the lagrangian $c_{W'}$ is the coefficient of the following operator:
\be
{\cal O}_{W'}=\frac{1}2 c_{W'}  g^2 H^\dagger H\, W_\mu^{\prime +} W^{\prime -\, \mu} \ .
\ee
For the SM $W$ boson we have $c_{W}=1$.
Using the exact one-loop form factors in Eqs.~(\ref{eq:loop1}) and (\ref{eq:loop2}), we define the enhancement factor over the SM diphoton width:
\be
R_{\gamma\gamma} = \left| 1+ c_{W'} \frac{m_W^2}{m_{W'}^2}\frac{A_1(\tau_{W'})}{  A_1(\tau_W)+ N_c Q_t^2 \, A_{1/2}(\tau_t) }\right|^2\ .
\ee
In the limit $m_{W'}\to \infty$, the leading contribution from the $W'$ loop becomes
\be
c_{W'} \frac{m_W^2}{m_{W'}^2} A_1(\tau_{W'}) \rightarrow -7 \times c_{W'} \frac{m_W^2}{m_{W'}^2} \ ,
\ee
in accordance with Eq.~(\ref{eq:Lhgaga1}). From Fig.~\ref{fig1} we see that, for a positive $c_{W'}$, an enhancement by a factor of two is possible for $c_{W'} \agt 1$ and  $m_{W'}\agt 130$ GeV. We note in passing that the same enhancement can be achieved for a similar mass range  if  $c_{W'}\alt -5$, which  requires large couplings and some fine tuning between the two contributions to the $W'$ mass in Eq.~(\ref{mW'}) in order to get a light $W'$.  

%%%%%%%%%%%%%%%%%%%%%%%%%%%%%
%%%%%%%%%%%%%%%%%%%%%%%%%%%%%%%%  

Notice there is a correlation between the sign of $c_{W'}$ and the cancellation of, or the lack thereof, the quadratic divergence in the Higgs mass-squared due to the SM $W$ boson  \cite{Low:2009di}. For $c_{W'} > 0\, (< 0)$, the $W'$ boson adds to (cancels) the quadratic divergences induced by the $W$ boson, which in the SM partially offsets the dominant top quadratic divergences. 

A $W'$ boson with direct couplings to the SM quarks and leptons is severely constrained by direct searches at the Tevatron and the LHC. Assuming SM coupling strengths, the lower bound on the mass for decays into leptonic final states is in the multi-TeV region \cite{wprimesearch} while searches in the dijet resonances lead to a  weaker bound, at around 850 GeV \cite{Aaltonen:2008dn}. Thus the $W'$ boson giving rise to the enhancement in the diphoton cannot couple to the SM quarks and leptons directly. One possibility is to impose a new $Z_2$ parity in the same fashion as the KK-parity in universal extra-dimensions \cite{Appelquist:2000nn} and the T-parity in little Higgs theories \cite{Cheng:2003ju}.  For example, the bound on the $W'$ boson in little Higgs theories with T-parity from precision electroweak constraints is $\agt 280$ GeV \cite{Cheng:2003ju}. Below that mass additional particles would need to be present to cancel the $W'$ contribution to the $\rho$ parameter. We will see in the next section that the same $W'$ boson in the loop of the diphoton width would also modify the Higgs decay width in the $Z\gamma$ channel. Therefore,  if simultaneous measurements of $\gamma\gamma$ and $Z\gamma$ widths point to a light $W'$ boson as the underlying mechanism, it would definitely hint at additional structures and particles at the TeV scale.  Given current constraints on direct searches, such a $W'$ could  decay into dijet plus a missing particle which is the lightest parity-odd particle. A possible discovery mode in colliders would be pair-production of the $W'$ bosons decaying into four jets plus missing transverse energy, which has not been searched for at the LHC.

 \subsection{New charged scalars}
 
\begin{figure}[t]
\includegraphics[scale=0.5, angle=0]{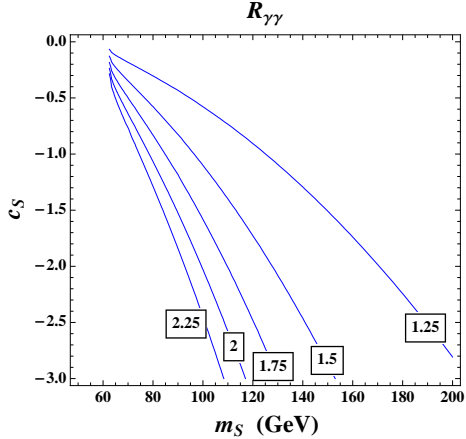}  
\caption{\label{fig2}{\em Contours of constant diphoton partial width, normalized to the SM value, shown as a function of $c_S=g_{hSS}/v$ and  the new  scalar mass.
}}
\end{figure}

We consider  one new scalar first, and by analogy to the $W'$ boson case,   we parametrize the mass of the new electrically charged scalar as
 \be
 \label{eq:onesmass}
 m_S^2= m_{S0}^2 + \frac12 c_S\, v^2 \  ,
 \ee
 where $m_{S0}$ is independent of $v$. The operator giving rise to $c_S$ is
 \be
 {\cal O}_S = c_S H^\dagger H\, |S|^2 \ ,
 \ee
 which results in $g_{hSS} = c_S v$.  Contrary to the $W'$ case, to get an enhancement, we would need to  assume $c_S<0$ so that the scalar contribution interferes constructively with the SM $W$ boson loop. The case of $c_S>0$ requires a scalar mass that is lighter than the case we discuss. Considering $Q_S=1$ as an example, the enhancement factor is 
 \be
 \label{eq:gagaS}
R_{\gamma\gamma} = \left| 1+  \frac{c_{S}}2 \frac{v^2}{m_{S}^2}\frac{A_0(\tau_{S})}{ A_1(\tau_W)+ N_c Q_t^2 \, A_{1/2}(\tau_t)}\right|^2\ ,
\ee
 For  $c_S \alt -2$, an enhancement of a factor of two is possible for $m_S\agt 100$ GeV, as can be seen in Fig.~\ref{fig2}.  For a heavier scalar mass, $m_S\agt 200$ GeV, a strong enhancement requires a very large $hSS$ coupling: $g_{hSS}/v= c_S\alt -10$.  A  negative $g_{hSS}$ coupling implies the following quartic couplings in the scalar potential:
 \be
 V(S, H) \supset - |c_S| |H^\dagger H| |S^\dagger S| + \frac{\lambda}{2} |H^\dagger H|^2 +\frac{ \lambda_S}{2} |S^\dagger S|^2 \ ,
 \ee
which could induces new charge breaking minima as well as problems with Higgs vacuum stability, if $|c_S|$ is large. A full analysis of these issues for a singlet scalar and a doublet scalar can be found in Ref.~\cite{Barger:2008jx}. For example, the condition that the scalar potential is bounded from below requires
\be
|c_S|^2 < {\lambda_S \lambda} . 
\ee
Since the Higgs quartic coupling is fixed by the Higgs mass ($\lambda \simeq 0.25$), a large value  of $|c_S|$ demands very large values of $\lambda_S$. Therefore,  values of $|c_S|$ larger than a few units would  either lead to strong couplings or be in conflict with vacuum stability, unless additional contributions are present to stabilize the potential.

One could achieve the strong enhancement with a heavier scalar mass  by assuming a larger charge, like the doubly charged scalar in an electroweak triplet, and/or a large number of degrees of freedom, $\tilde{N}_{c,S}$, associated with a ``dark color'' charge different from the $SU(3)_c$ one. Since the contribution of the charged scalar to the amplitude grows with $\tilde{N}_{c,S} Q_S^2/m_S^2$ parametrically,  then one can obtain the same enhancement for larger masses by scaling up $\tilde{N}_{c,S}$ and/or $Q_S$, and the scaling goes like 
 \begin{equation}
 m_S^2 \simeq \sqrt{\tilde{N}_{c,S}} \; |Q_S| \left(m_S^2\right)_{\tilde{N}_c=Q_S = 1}.
 \end{equation}
 Still, unless unnatural values of the charges or colors are assumed,  in order to get a significant enhancement of the diphoton rate, the new scalars must have masses below the weak scale. One could also use a large value of  $\tilde{N}_{c,S}$ to achieve a significant enhancement with a positive $c_S$, in order to avoid the vacuum instability associated with a  large, negative $c_S$.  For a factor of two enhancement in the diphoton width that can be achieved by a particular choice of  $(-|c_S|, m_S)$, $\tilde{N}_{c,S}\sim 6$ is needed for the same enhancement from $(+|c_S|, m_S)$. Even larger $\tilde{N}_{c,S}$ is necessary if the measured increase in the diphoton width would become smaller.
 
%%%%%%%%%%%%%%%%%%%%%%%%%%%%%
%%%%%%%%%%%%%%%%%%%%%%%%%%%%%%%%  

A natural way of obtaining negative couplings of the scalars to the Higgs boson is via scalar mixing, which can be seen easily from the general arguments presented in Section \ref{sect:II}. Basically it boils down to the observation that the Higgs coupling to photons is controlled by the determinant of the mass-squared matrix and the mass mixing always reduces the determinant.  It is also possible to see the same effect by going directly into the mass eigenbasis in the presence of mixing. We will see that in the mass eigenbasis the lighter mass eigenstate could obtain an ``effective'' $g_{hSS}$ coupling which is negative. The canonical example is the mixing between an electroweak doublet scalar and a singlet scalar carrying the quantum numbers of the left-handed and right-handed leptons, respectively, which appears in supersymmetry~(see, for example, Refs.~\cite{Carena:2011aa} and \cite{Carena:2012gp}). In this case the mass mixing occurs only after the electroweak symmetry breaking and requires an insertion of the Higgs VEV, which implies that the mass mixing not only affects the mass eigenvalues, but also directly the coupling of the mass eigenstates to the Higgs boson. If the two charged scalars have the same electroweak quantum number and the mixing does not go through a Higgs insertion, then the Higgs coupling to the mass eigenstates depends on the mixing parameter only implicitly through the mixing angles between the gauge and mass eigenbasis, and would not have a big effect on the partial width. Therefore, in the following we focus on the canonical example of mixing between a doublet scalar and a singlet scalar. 
 
Denoting the two charged scalars in the gauge basis by $S_L$ and $S_R$,  one can write down the general mass-squared matrix,
 \be
 \label{eq:scalarmass}
{\cal M}_S^2 = \left( \begin{array}{cc}
           \tilde{m}_L(v)^2 &  \frac{1}{\sqrt{2}}v X_S \\
           \frac{1}{\sqrt{2}}v X_S & \tilde{m}_R(v)^2 
           \end{array} \right)\ ,
\ee
where $X_S$ is a dimensionful parameter characterizing the mass mixing. The mass matrix can be diagonalized by a $2\times 2$ rotation matrix,
\be
\label{eq:rotmat}
{\cal R}_S = \left( \begin{array}{cc}
           c_{\theta_S} &   s_{\theta_S} \\
           -s_{\theta_S} &  c_{\theta_S}
           \end{array} \right)\ ,
\ee
such that the mixing angle and mass eigenvalues are
\bea
&& m_{S_{1,2}}^2 = \frac12 \left[ \tilde{m}_L^2+\tilde{m}_R^2 \mp \sqrt{(\tilde{m}_L^2-\tilde{m}_R^2)^2+2 v^2 X_S^2}\right] \ ,\\
&&s_{2\theta_S} = -\frac{\sqrt{2} v X_S}{m_{S_1}^2-m_{S_2}^2} \ ,\qquad c_{2\theta_S} = \frac{\tilde{m}_L^2-\tilde{m}_R^2}{m_{S_1}^2-m_{S_2}^2} \ , 
\eea
We see that, when the off-diagonal term is large $v X_S\gg (\tilde{m}_L^2-\tilde{m}_R^2)$, the mixing angle is maximal: $s_{2\theta_S}\approx 1$ and $c_{2\theta_S}\approx 0$. Notice that our notation is such that $S_1$ and $S_2$ are the lighter and the heavier mass eigenstates, respectively.

The effect of the mass mixing may be understood in two ways. First, by computing the properties of the determinant, as done in the previous section, and second by computing the dominant effects provided by the lightest scalar. We define, similar to Eq.~(\ref{eq:onesmass}),
\be
\tilde{m}_L^2= \tilde{m}_{L0}^2 + \frac12 {c}_{L} v^2 \ , \qquad  \tilde{m}_R^2= \tilde{m}_{R0}^2 + \frac12 {c}_{R} v^2 \ ,
\ee
From Eq.~(\ref{eq:generalmix}) we get 
\begin{equation}
\frac{\partial \log\left(\det {\cal M}_S^2 \right)}{\partial v} \simeq   v \frac{(m_{L0}^2 + \frac12 {c}_L v^2) {c}_R + (m_{R0}^2 + \frac12 {c}_R v^2) {c}_L  - X_S^2 }{m_{S_1}^2 m_{S_2}^2} . 
\end{equation}
Since the denominator is positive this shows that a constructive interference with the $W^\pm$ gauge boson contribution demands either negative coefficients ${c}_{L,R}$ or
a large mixing contribution from $X_S$.  Alternatively, we can compute the effective $g_{hS_iS_i}$ couplings in the $(S_1,S_2)$  eigenbasis by using Eq.~(\ref{eq:generalcoup}),
\bea
  \label{photocoup}
g_{hS_1S_1} &=&   {c}_{+} v + c_{2\theta_{S}} {c}_{-} v - \frac{1}{\sqrt{2}}s_{2\theta_{S}} X_S \ , \\
g_{hS_2S_2} &=&  {c}_{+} v - c_{2\theta_{S}}  {c}_{-} v + \frac{1}{\sqrt{2}}s_{2\theta_{S}} X_S \ ,
\eea
where ${c}_{\pm} = ({c}_{L}\pm {c}_{R})/2$. The effective $c_{S_i}$ are defined as
\be
c_{S_i} = \frac{g_{hS_iS_i}}{v} \ , \quad i=1,2 \ .
\ee
Note that the sign of the term proportional to the off-diagonal $X_S$ term in the $g_{hS_iS_i}$ coupling is always negative for the lighter mass eigenstate, which is why the mass eigenvalue is smaller. Therefore, when ${c}_{L,R}$ are both negative, the mixing parameter $X_S$ further enhances the $g_{hS_1S_1}$ coupling. Even when both ${c}_{L,R}$ are positive, when $X_S$ is large and the mixing maximal,  the sign of $g_{hS_1S_1}$ coupling could be flipped from positive to negative, in which case $S_1$ interferes constructively with the SM $W$ boson as if it acquired a negative effective coupling $c_S$.\footnote{It is worth pointing out that, since the photon coupling is ``vector-like,'' the $hS_1 S_2$ coupling is not involved in the diphoton width; only $hS_i S_i$ couplings enter.}  So, by focusing on  $S_1$ as the dominant contribution to the diphoton decay width, we obtain similar conclusions to the ones obtained by the analysis of the scalar determinant above. 
 
 %%%%%%%%%%%%%%%%%%%%%%%%%%%%%
\begin{figure}[t]
\includegraphics[scale=0.43, angle=0]{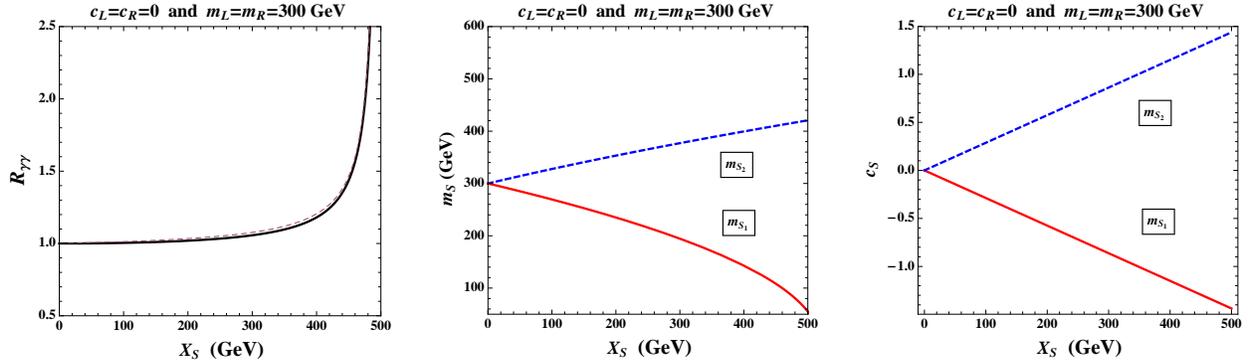}  
\caption{{\em Left panel: Diphoton partial width normalized to the SM as a function of the mixing parameter between the two charged scalars. The solid (dashed) line in the $R_{\gamma\gamma}$ plots includes both (only the lightest) mass eigenstates. They are almost on top of each other since the contribution from the heavy mass eigenstate is tiny. Middle panel: Mass of the lightest (solid, red line) and heaviest (dashed, blue line) scalar mass eigenstates as a function of the mixing parameter. Right panel: Effective couplings of the lightest (solid, red line) and heaviest (dashed, blue line) scalar mass eigenstates as a function of the mixing parameter. }}
\label{fig2-1}
\end{figure}
%%%%%%%%%%%%%%%%%%%%%%%%%%%%%%%%  

As an example, in Fig.~\ref{fig2-1} we show the enhancements in the diphoton width as a function of the mixing parameter $X_S$ for the following scenario:
\be
 {c}_L={c}_R=0 \quad {\rm and} \quad m_L=m_R= 300 \ \ {\rm GeV} \nonumber \ .\\
 \ee
The solid and dashed lines in the $R_{\gamma\gamma}$ plots are for including both mass eigenstates and only the lightest mass eigenstate, respectively. We see that the contribution from the heavier $S_2$ is negligible, as the dashed line is right on top of the solid line in the left panel of Fig.~\ref{fig2-1},  which implies the enhancement is entirely due to the lighter eigenstate $S_1$.  An enhancement by a factor of 1.5 is possible for $X_S \agt 450$ GeV, for which $m_{S_1}\agt 120$ GeV and $c_{S_1} \alt -1.3$. 

In general, larger values of  $m_L$ and $m_R$ require  larger  values of the mixing parameter $X_S$ in order to get a significant enhancement.  Parametrically the critical value of $X_S$ for a large enhancement grows  with  $m_L m_R$, which is the positive contribution to the determinant of the mass-squared matrix.  It is easy to see that large values of $X_S \gg v$ induce the presence of charge breaking minima, deeper than the electroweak one. (See, for example, Ref.~\cite{Hisano:2010re}.) Hence, scenarios with $X_S \simgt 1$~TeV require additional new physics at the weak scale to stabilize the vacuum. In all realistic cases, a large enhancement of the Higgs diphoton width demands masses of scalars below the weak scale. 

Light charged scalars have been searched for at colliders. For example, LEP put a lower bound on the mass of sleptons in supersymmetry  that is of the order of 100 GeV \cite{lepstau}. Similar to the $W'$ case, one could  postulate a new $Z_2$ parity carried by the new scalar, much like the R parity carried by the sleptons. While we have not specified a detailed production and decay mechanism of the charged scalar under consideration, we note that a somewhat large coupling to the Higgs boson is necessary in order to have a scalar mass heavier than the lower bound on the slepton mass, as can be seen in Figs.~\ref{fig2} and \ref{fig2-1}.  Indeed Ref.~\cite{Carena:2011aa}, found that, in MSSM, a significant enhancement in the diphoton channel is possible only when the stau is very light, close to its direct search limit. Possible stau search strategies at the LHC were subsequently discussed in Ref.~\cite{Carena:2012gp}.
 
 \subsection{New charged leptons}
 
\begin{figure}[t]
\includegraphics[scale=0.5, angle=0]{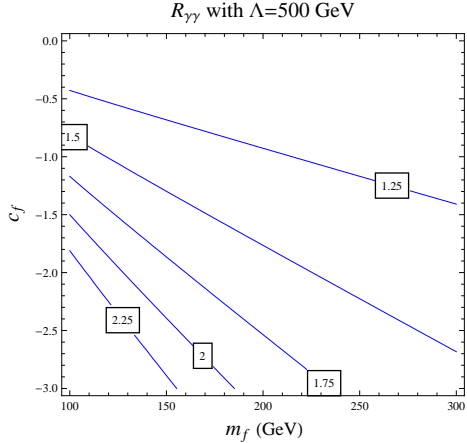}  
\caption{\label{fig3}{\em Contours of constant diphoton partial width, normalized to the SM value, shown as a function of $c_f=g_{h\bar{f}f} \Lambda/v$ and  the new  vector-like lepton mass
}}
\end{figure}

 Here the leptons are defined as any charged fermion carrying no color. The discussion is very similar to the scalar case.   Again we start with one new vector-like pair of charged leptons,  whose  mass term is written as
 \be
 m_f = m_{f0} + c_f \frac{v^2}{2\Lambda} \ ,
 \ee
 where $\Lambda$ is a dimensionful parameter. Since the lepton is vector-like and has a Dirac mass term,  $c_f$ can only originate from the dimension-five operator,
 \be
 \label{eq:fermiondim5}
 {\cal O}_f = \frac{c_f}{\Lambda} H^\dagger H \bar{f} f \ ,
 \ee
 giving rise to $g_{hf\bar{f}}=c_f v/\Lambda$.  Then the enhancement factor is
 \be
 \label{eq:gagaf}
R_{\gamma\gamma}= \left| 1+ c_{f} \frac{v^2}{\Lambda m_f}\frac{A_{1/2}(\tau_{f})}{  A_1(\tau_W)+ N_c Q_t^2 \, A_{1/2}(\tau_t) }\right|^2\ ,
\ee
where we have assumed $Q_f=N_{c,f}=1$. Similar to the scalar case we focus on $c_f<0$. In Fig.~\ref{fig3} we see a factor of two increase in the diphoton width requires $c_f \alt -2$ for $m_f \agt 140$ GeV with $\Lambda=500$ GeV.  
 
 %%%%%%%%%%%%%%%%%%%%%%%%%%%%%
%%%%%%%%%%%%%%%%%%%%%%%%%%%%%%%%  
 
Next we discuss the possibility of fermion mass mixing. Reasonings similar to the discussion of scalar mass mixing in the previous subsection lead us to introduce a vector-like pair of charged fermions $(\ell_4,\ell_4^c)$ carrying the same quantum number as the left-handed charged leptons, as well as a vector-like pair of fermions $(L_4,L_4^c)$ with the same quantum number as the right-handed charged leptons. The mass mixing is  then induced by Yukawa-like couplings between $(\ell_4, L_4^c)$ and $(L_4, \ell_4^c)$ after electroweak symmetry breaking. We do not wish to introduce a fourth-generation-like leptons, which would always interfere destructively with the SM $W$ boson loop, much like the top quark does, and an overall enhancement is difficult to obtain.

 The fermion mass matrix is written as follows 
 \be
(\ell_4^{c}, L_4^{c}){\cal M}_f 
\left( \begin{array}{c}
                    \ell_4 \\
                    L_4
                    \end{array} \right)
 =(\ell_4^{c}, L_4^{c}) \left( \begin{array}{cc}
           m_{\ell_4}(v) &   Y_f v \\
       Y_f  v & m_{L_4}(v) 
           \end{array} \right)
        \left( \begin{array}{c}
                    \ell_4 \\
                    L_4
                    \end{array} \right)\ ,
\ee
where for simplicity we have assumed a single Yukawa coupling $Y_f$ controlling both the $(L_4^c,\ell_4)$ and $(\ell_4^c,L_4)$ mass mixings, and 
\be
m_{\ell_4}(v) = m_{\ell_40} + {c}_{\ell_4} \frac{v^2}{2\Lambda} \ , \qquad m_{L_4}(v) = m_{L_40} + {c}_{L_4} \frac{v^2}{2\Lambda} \ .
\ee
For simplicity, we shall assume that all coefficients are real. 
To solve for the mass eigenvalues, we diagonalize the mass matrix-squared,
\be
{\cal M}_f^\dagger {\cal M}_f = 
\left( \begin{array}{cc}
           m_{\ell_4}^2(v) + Y_f^2 v^2& (m_{\ell_4} +m_{L_4})  Y_f v\\
           (m_{\ell_4} +m_{L_4})  Y_f  v & m_{L_4}^2(v) + Y_f^2 v^2
           \end{array} \right)
\ee           
which is in a form similar to the scalar case in Eq.~(\ref{eq:scalarmass}). 

The conclusion from the analysis of the determinant of the mass matrix-squared is similar to the scalar case, and a constructive interference with the SM $W$ loop could be obtained for negative ${c}_{\ell_4}$ and ${c}_{L_4}$ and/or large mass mixing. In the mass eigenbasis the mass eigenvalues and mixing angles are 
\bea
&& m_{f_{1,2}}^2 = \frac12 \left[ m_{\ell_4}^2+m_{L_4}^2+2Y_f^2 v^2 \mp \sqrt{(m_{\ell_4}^2-m_{L_4}^2)^2+4 (m_{\ell_4} +m_{L_4})^2  Y_f^2 v^2}\ \right] \ ,\\
&&s_{2\theta_f} = -\frac{2(m_{\ell_4} +m_{L_4}) Y_f v}{m_{f_1}^2-m_{f_2}^2} \ ,\qquad c_{2\theta_f} = \frac{m_{\ell_4}^2-m_{L_4}^2}{m_{f_1}^2-m_{f_2}^2} \ .
\eea
  %%%%%%%%%%%%%%%%%%%%%%%%%%%%%
\begin{figure}[t]
\includegraphics[scale=0.4, angle=0]{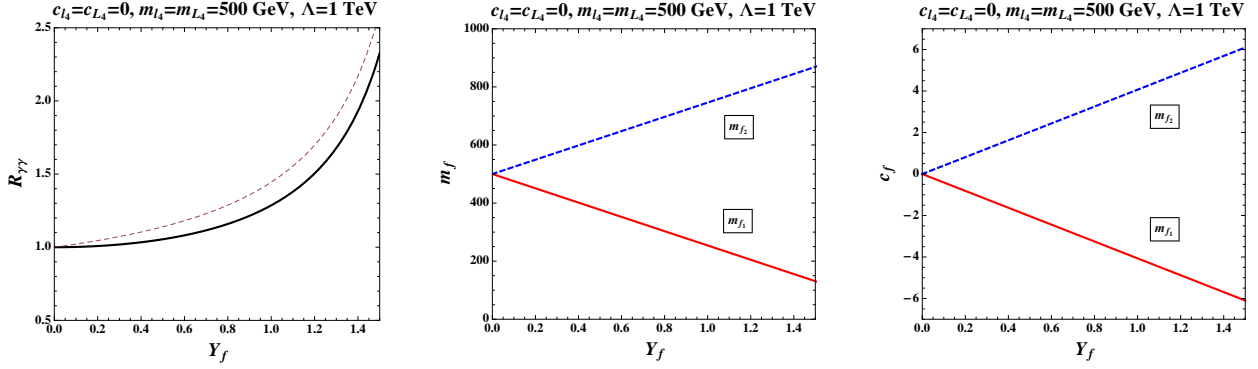}  
\caption{\label{fig3-1}{\em 
Left panel: Diphoton partial width normalized to the SM as a function of the mixing parameter between the two vector-like leptons. The solid (dashed) line in the $R_{\gamma\gamma}$ plots includes both (only the lightest) mass eigenstates.  Middle panel: Mass of the lightest (solid, red line) and heaviest (dashed, blue line) vector-like lepton mass eigenstates as a function of the  mixing parameter. Right panel: Effective couplings of the lightest (solid, red line) and heaviest (dashed, blue line) vector-like lepton mass eigenstates to the Higgs bosons as a function of the mixing parameter. }}
\end{figure}%
%%%%%%%%%%%%%%%%%%%%%%%%%%%%%%%%  
The $g_{hf_i\bar{f}_i}$ couplings are again obtained from Eq.~(\ref{eq:generalcoup}). In Fig.~\ref{fig3-1} we show the enhancements in the diphoton width as a function of the mixing parameter $Y_f$ for the following scenario:
\be
  c_{\ell_4}=c_{L_4}=0 \ ,  \quad m_{\ell_4 0}=m_{L_4 0} =500 \ \ {\rm GeV}   \ , \quad  {\rm and} \quad  \Lambda= 1 \ \ {\rm TeV} \nonumber  \ .
 \ee
Again the solid and dashed lines in the $R_{\gamma\gamma}$ plot are for including both mass eigenstates and only the lightest mass eigenstate, respectively. We see that the contribution from the heavier $f_2$ is small, and has the effect of reducing the diphoton rate, since the effective coupling is positive.  This implies the enhancement is largely determined by the lighter eigenstate $f_1$.  This feature is already present in the scalar case in Fig.~\ref{fig2-1}, although the contribution of the heaviest scalar is much harder to discern due to the smallness of the effect. The lighter state $f_1$ has a negative effective coupling to the Higgs and could strongly enhance the diphoton width. An enhancement by a factor of 1.5 is possible for $Y_f \agt 1.2$, for which $m_{f_1}\alt 200$ GeV and $c_{f_1} \alt -4$.  

The fact that, contrary to the scalar case,  the heavier mass eigenstate does make a non-negligible contribution is due to a number of factors. First of all, from the right panel in Fig.~\ref{fig3-1} we see that the effective couplings, $c_{f_i}, i=1,2$, are of equal magnitude, but opposite in sign, in the benchmark scenario we considered. Also the loop function $A_{1/2}$ remains an order unity factor in the mass range we studied. Then from Eq.~(\ref{eq:gagaf}) we see that the contribution from the heavy state is suppressed by $m_{f_1}/m_{f_2}$ relative to that from the light state.\footnote{This pattern of suppression is different in  the case of the top partners in little Higgs theories \cite{ArkaniHamed:2001nc}. In these models the Dirac mass term of the fermion is related to the dimension-five operator in Eq.~(\ref{eq:fermiondim5}) due to the non-linearly realized global symmetry acting on the Higgs boson. The suppression in the contribution of the heavy top partner $T$ relative to that from the SM top quark $t$ turns out to be $m_{t}^2/m_{T}^2$. } On the other hand, for similar reasons, in the scalar case the suppression of the heavier state contribution relative to the lighter state one is proportional to  $m^2_{S_1}/m^2_{S_2}$,  as seen from Eq.~(\ref{eq:gagaS}), and therefore decouples faster than in the fermion case.
 
It is worth noting that  fermions with a large Yukawa coupling could spoil the stability of the Higgs potential, which is well-known in the context of the fourth-generation models \cite{Sher:1988mj}. Although we are considering vector-like fermions, whose effects should decouple, the fermion masses necessary for a significant enhancement in the diphoton channel are still quite light, of the order of a few hundred GeV, and their effects may not decouple fast enough. More concretely,  large values of the Yukawa couplings tend to induce a reduction of the Higgs quartic coupling at high energies, leading to potential instabilities in the Higgs potential. For values of $Y_f$ of order one or larger, such instabilities occur at the TeV scale. Therefore, as in the scalar case, a large increase of the diphoton rate must be associated with new physics at the TeV scale, beyond the one leading to the diphoton rate enhancement, to stabilize the Higgs potential. 

Searches for light charged fermions are again performed in, for example, the context of charginos in supersymmetry at the LEP. A lower bound on the chargino mass is again of the order of 100 GeV \cite{lepschargino}. Again one could assume a new $Z_2$ parity carried by the new fermion like the R parity carried by the charginos. From Figs.~\ref{fig3} and \ref{fig3-1} we see that, similar to the scalar case, a somewhat large coupling to the Higgs is required to have a light charged fermion heavier than the chargino mass bound.
 
 \section{Correlating the Higgs $\gamma\gamma$ width with the $Z\gamma$ width}
 \label{sect:IV}
 
 Apart from the diphoton coupling, the Higgs coupling to $Z \gamma$ is also induced at the loop level by the same particles running in the loop, due to the electroweak gauge symmetry. One can therefore expect a correlation between an enhancement in the diphoton width with a shift in the $Z\gamma$ width.  The SM contributions to the $Z\gamma$ width is given by \cite{Cahn:1978nz}:
 \be
 \Gamma(h\to Z\gamma)= \frac{G_F^2 m_W^2 \alpha}{64\pi^4}  m_h^3 \left(1-\frac{m_Z^2}{m_h^2}\right)^3 \left| {\cal A}_{SM} \right|^2 \ ,
 \ee
 where
 \be
{\cal A}_{SM} = \cos \theta_w A_1(\tau_W,\lambda_W) + N_c \frac{Q_t (2T_3^{(t)}-4 Q_t s_w^2)}{c_w} A_{1/2}(\tau_t,\lambda_t) \
\ee
with  $\tau_i=4m_i^2/m_h^2$, $\lambda_i=4m_i^2/m_Z^2$, and $T_3^{(t)}=1/2$ is the weak isospin of the top quark whereas  $Q_t = 2/3$ is its electric charge in units of $|e|$. More generally, including contributions from new charged particles that do not carry any color charge, we can write
 \be
 \label{eq:Zgageneral}
 \Gamma(h\to Z\gamma)= \frac{\alpha^2}{512\pi^3} m_h^3 \left(1-\frac{m_Z^2}{m_h^2}\right)^3 \left|\frac{2}{v}\frac{{\cal A}_{SM}}{\sin\theta_w} +  {\cal A} \right|^2 \ ,
 \ee
 where
 \bea
 \label{eq:Zga1}
{\cal A}&=&\frac{g_{hW'W'}}{m_{W'}^2} g_{ZW'W'} A_1(\tau_{W'},\lambda_{W'})  + \tilde{N}_{c,f}  \frac{2g_{hf\bar{f}}}{m_f} (2Q_f)\;(g_{Z\ell\ell}+g_{Zrr}) A_{1/2}(\tau_f,\lambda_f) \nonumber \\
&& \qquad -  \tilde{N}_{c,S} \frac{2g_{hSS}}{m_S^2} Q_{S} \; g_{ZSS} A_0(\tau_S,\lambda_S) \ .
 \eea
 In the above $g_{hii}$ and $g_{Zii}$ are the Higgs and $Z$ couplings to a pair of the $i$ particle, respectively, and  we have considered the $Z$ coupling to left-handed and right-handed fermions separately.  Moreover, we have also included the possibility that the fermions and scalars  have additional $\lq\lq$dark color" degrees of freedom $\tilde{N}_{c,f}$ and $\tilde{N}_{c,S}$, respectively, different from the $SU(3)_c$ ones. 
 A consistency check of the scalar contribution in Eq.~(\ref{eq:Zga1}) is given by the requirement that, in the limit $m_Z\to 0$, the scalar contribution reduces to two times that in the diphoton amplitude \cite{Chen:2013vi}. The $Z$ boson couplings to fermions are,  
\be 
 g_{Z\ell\ell}=\frac{1}{s_w c_w}(T^{(\ell)}_3-Q s_w^2) \ , \qquad {\rm and} \qquad g_{Zrr}= \frac{1}{s_w c_w}(T_3^{(r)}-Q s_w^2)\ ,
 \ee
 where $T_3^{(\ell)}$ and $T_3^{(r)}$ are the weak isospin of the left-handed and right-handed fermions, respectively, and the electric charge $Q$ is in unit of $|e|$.   Notice that our definition of $ A_1(\tau_w,\lambda_w)$ differs from that in Ref.~\cite{Gunion:1989we} by a factor of $\cot \theta_w$. The modification in the partial decay width of the Higgs in the $Z\gamma$ channel is then expressed in terms of
 \be
 R_{Z\gamma} = \left| 1 + \frac{{\cal A}}{(2/v)({\cal A}_{SM}/s_w)} \right|^2 \ .
 \ee
When the mass eigenstates are admixtures of particles with different isospin quantum numbers, there are diagrams that contain two different mass eigenstates in the loop. However, Eqs.~(\ref{eq:Zgageneral}) and (\ref{eq:Zga1}) describe only the contributions from loop diagrams containing the same mass eigenstate. We will argue later that, in the region of parameter space we are interested, the contribution from mixed diagrams where different mass eigenstates run in the loop is in general suppressed compared to the diagram containing only the lightest mass eigenstate.

It is worth pointing out that, unlike in the $\gamma\gamma$ channel where only the electric charge of the loop particle enters,  the amplitude in the $Z\gamma$ channel now involves the coupling of the loop particle to the SM $Z$ boson, which in turn depends on the $SU(2)_L \times U(1)_Y$ quantum number. Therefore, simultaneous measurements of the decay widths in the $\gamma\gamma$ and $Z\gamma$ channels would probe the weak isospin charge and the electric charge of the new particles running in the loop.

Below we will consider the modifications in the $Z\gamma$ channel first assuming there is only a single new particle inducing the enhancement in the diphoton channel, and then proceed to analyze the possibilities of mass mixing among new particles.

\subsection{No Mass Mixing}

For the $W'$ scenario, we assume that the $W'$ is the $T^3=\pm 1$ component of an electroweak triplet and therefore the $g_{ZW'W'}$ coupling is fixed to be the same as $g_{ZWW}$ due to the gauge invariance, 
\be
g_{ZW'W'} = g_{ZWW} = \cot \theta_w \ .
\label{eq:tgc}
\ee
The scalar and fermion cases, instead, depend on the specific electroweak quantum numbers. We consider two benchmarks where the scalars/fermions are $SU(2)_L$ singlets and doublets, respectively,
\bea
\label{eq:zgabench1}
{\rm (1)}&:& g^{(1)}_{ZSS} =\frac{1}{s_w c_w}( -Q_S s_w^2) \ ,\qquad \quad\ \  g^{(1)}_{Z\ell\ell} = g^{(1)}_{Zrr} = \frac{1}{s_w c_w}(-Q_f s_w^2)  \ , \\
{\rm (2)}&:& g^{(2)}_{ZSS} =\frac{1}{s_w c_w}( -\frac12 -Q_S s_w^2) , \qquad g^{(2)}_{Z\ell\ell} =g^{(2)}_{Zrr} = \frac{1}{s_w c_w}\left(-\frac12 -Q_f s_w^2\right) 
\label{eq:zgabench2}
 \eea
 with $Q_S=Q_f=-1$.
%%%%%%%%%%%%%%%%%%%%%%%%%%%%%
\begin{figure}[t]
\includegraphics[scale=0.52, angle=0]{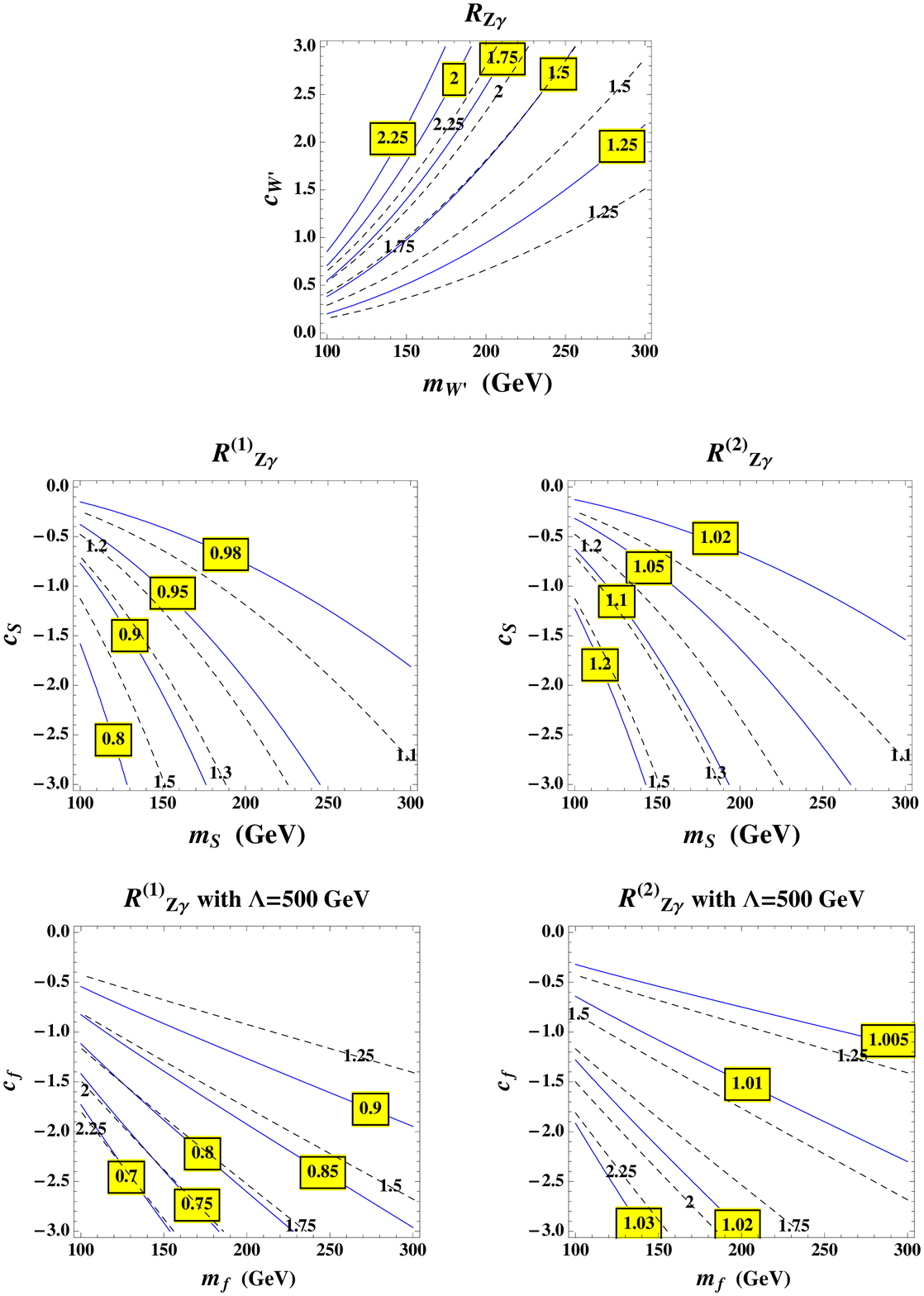}  
\caption{\label{fig6}{\em Enhancement in the $Z\gamma$ partial width due to a new particle. We have overlaid the changes in the diphoton width in the corresponding choices of parameters, which are shown as dashed curves. In the scalar and fermion cases we consider two benchmark scenarios, (1) and (2), as indicated in Eqs.~(\ref{eq:zgabench1}) and (\ref{eq:zgabench2}).
}}
\end{figure}
%%%%%%%%%%%%%%%%%%%%%%%%%%%%%%%%  
  
 In Fig.~\ref{fig6} we show the modifications in the $Z\gamma$ partial width of the Higgs boson and the corresponding enhancements in the diphoton width. In the $W'$ scenario we observe large enhancement in the $Z\gamma$ channel for $c_{W'}>0$. If $c_{W'}<0$, instead, one would obtain a significant reduction in this channel. Due to the small couplings of the charged particles to the $Z$ gauge boson, Benchmark 1 yields  moderate reductions in $Z\gamma$ channel both the scalar and the fermion cases, while the pattern of deviations  in Benchmark 2  is reversed between the scalar and the fermion cases with respect to Benchmark 1.

 \subsection{With Mass Mixing}

 %%%%%%%%%%%%%%%%%%%%%%%%%%%%%
\begin{figure}[t]
\includegraphics[scale=0.45, angle=0]{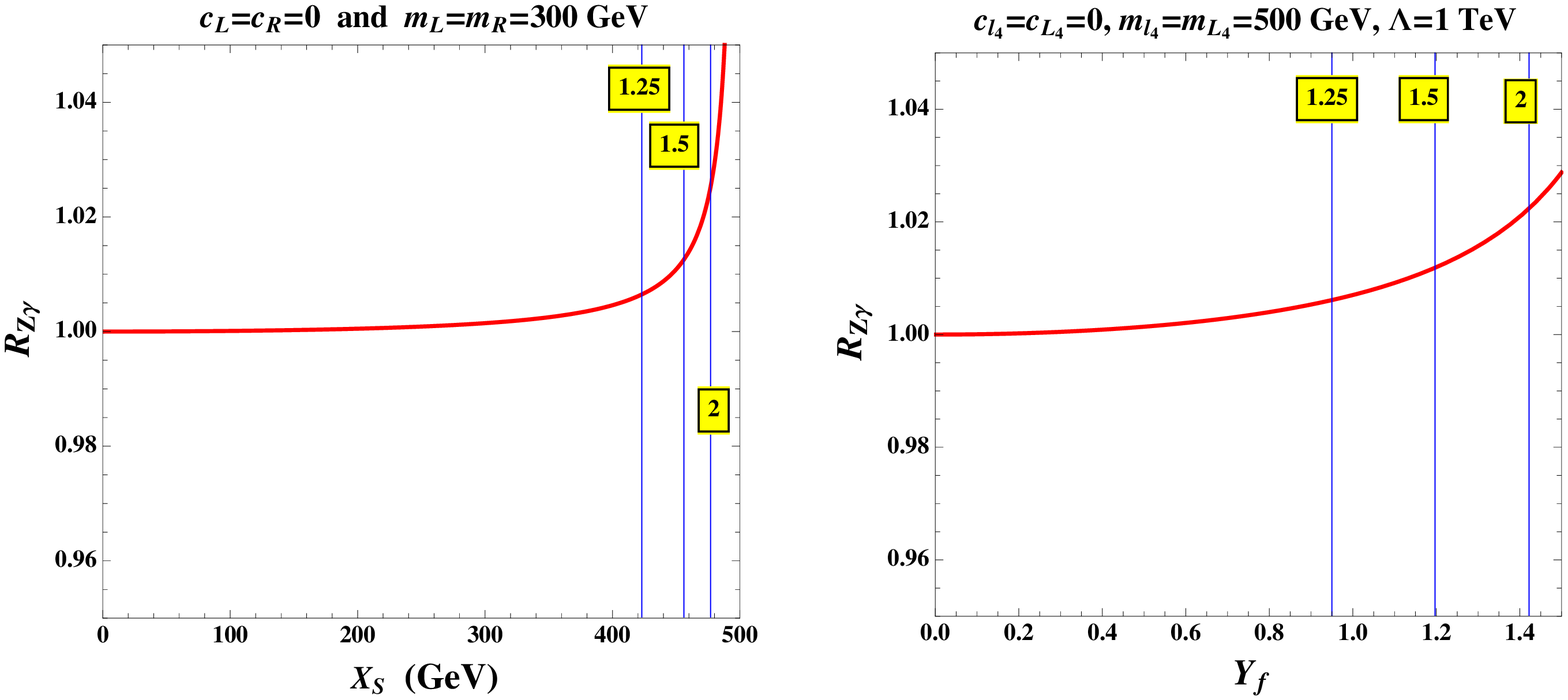}  
\caption{\label{fig7}{\em Enhancement in the $Z\gamma$ partial width due to mass mixings. The left panel is for the scalar mass mixing while the right panel is for the fermion mass mixing. The mixing parameters necessary to induce large enhancements in the diphoton widths are shown in blue (vertical) lines.
}}
\end{figure}
%%%%%%%%%%%%%%%%%%%%%%%%%%%%%%%%  
 
 When there is mass mixing between the new particles, the coupling of mass eigenstates to the SM $Z$ boson depends on the mixing angle. In the scalar case $S_L$ is an $SU(2)_L$ doublet while $S_R$ is assumed to be a singlet. The resulting couplings are
   \be
   \label{zsscoupling}
{\cal R}^T
    \left( \begin{array}{cc}
           T_3^{L} -Q_S s_w^2 &  0 \\
          0 &  - Q_S s_w^2 
           \end{array} \right) {\cal R}      =   
       \left( \begin{array}{cc}
           T_3^{L} c_{\theta_S}^2 - Q_S s_w^2    & s_{2\theta_{S}} T_3^{L}/2 \\
         s_{2\theta_{S}} T_3^{L}/2 &  T_3^{L} s_{\theta_S}^2 -Q_S s_w^2  
           \end{array} \right)\ ,
\ee
where ${\cal R}$ is the rotation matrix from flavor to mass eigenbasis defined in Eq.~(\ref{eq:rotmat}). Contrary to the coupling to photons, there is now an off-diagonal coupling $g_{ZS_1S_2}$ to the $Z$ boson, which, together with the off-diagonal coupling to the Higgs boson,
\be
\label{eq:ghs1s2}
g_{hS_1S_2}=s_{2\theta_{S}} {c}_{-} v+ \frac{1}{\sqrt{2}}c_{2\theta_{S}} X_S \ ,
\ee
would give rise to a mixed one-loop diagram with both the heavy and the light mass eigenstates in the loop. While such a contribution has been evaluated in Ref.~\cite{Djouadi:1996yq}, the analytic form of the loop functions are much more complicated due to the presence of two mass scales, and will not be reproduced here. Nevertheless, notice that the $Z$ off-diagonal coupling in Eq.~(\ref{zsscoupling}) is multiplied by $s_{2\theta}$ while the off-diagonal Higgs couplings in Eq.~(\ref{eq:ghs1s2}) are multiplied by either $s_{2\theta}$ or $c_{2\theta}$.  The large Higgs off-diagonal contribution, proportional to $X_S$, is proportional to $c_{2\theta}$, and hence the  dominant contribution to the mixed scalar diagram is proportional to $s_{4\theta}$.  This observation suggests that the mixed diagram is suppressed in the large mixing limit where $s_{2\theta} \approx 1$. There is also the additional suppression from the loop function containing one heavy mass scale, relative to the contribution from the lightest mass eigenstate.  Since  a large contribution to the diphoton rate can only come from the case in which the mixing between the two scalars is large, one would expect a small contribution from the mixed diagram to the Higgs $Z\gamma$ decay width in this region of parameter space. 
  
In Fig.~\ref{fig7} we present the change in the $Z\gamma$ partial width in the mass mixing cases, concentrating on  the region of parameters giving rise to large enhancements in the diphoton channel. The mixing parameters that are necessary to induce large enhancements in the $\gamma\gamma$ channel are shown in vertical lines.  In general the modification to the  $Z\gamma$ partial width is insignificant,  at most a  5\% deviation from the SM expectation. This is to be expected not only because  the $Z$  couplings are suppressed compared to the electromagnetic couplings,  but also due to the fact that the mass eigenstates in these scenarios are mixtures of the charged component in the $SU(2)_L$ doublet and the singlet particle. As can be observed in Fig.~\ref{fig6}, the doublet and the singlet have opposite trends in terms of the interference pattern with the SM amplitudes. Therefore, the two effects tend to cancel each other in the mixing case, resulting in small deviations. However, it is possible that with more exotic choices of electroweak quantum numbers we could get a larger effect in the $Z\gamma$ channel, just like in the $W'$ case, which has the quantum numbers of an electroweak triplet.
 
 \section{Conclusion}
 \label{sect:V}
 
 In this work, we have analyzed the possibility that the observed Higgs diphoton decay width is a result of new physics, and we discuss the properties that the new particles should fulfill in order to explain such an enhancement. We have concentrated on the cases of new charged particles of spin zero, spin one-half, and spin one.  In general, a large enhancement of the Higgs diphoton decay width may only be obtained for particles with masses of  the order of a few hundred GeV. Depending on the size of the coupling of these new particles with the Higgs boson,  considerations from precision electroweak measurements and vacuum stability may hint at the existence of additional new particles at the TeV scale. In addition, to avoid constraints from direct searches for light charged particles that affect the diphoton partial width, one could postulate a new $Z_2$ parity carried by these new particles.  At the LHC, these particles could be accessible and pair-produced via electroweak processes, much like superpartners.

  In the spin zero case a constructive interference with the SM $W^\pm$ contribution demands either negative couplings to the Higgs or large mixing between, for instance, scalars transforming as doublets and singlets of the electroweak $SU(2)_L$ group. In the fermion case, we consider the case of vector-like fermions since chiral fermions tend to induce a reduction of the Higgs diphoton decay width. For vector-like fermions an enhancement can  again be obtained by either a negative coefficient of the dimension-five coupling of the fermion to the Higgs, or the presence of large mixing between species of different $SU(2)$ quantum numbers.  In the vector case, we have parametrized the vector coupling in unit of the SM $W^\pm$ coupling to the Higgs. In this case, positive couplings lead to constructive interference with the $W^\pm$ loop.

We have also studied the possible correlation with the $Z \gamma$ coupling.  In the fermion and scalar cases, we concentrated on $SU(2)_L$ singlet and doublet scenarios, and found the contribution to the $Z \gamma$ coupling to be significantly smaller than in the diphoton case. In the vector case the enhancement of the $\gamma \gamma$ Higgs decay width would be accompanied by a similar enhancement of the $Z \gamma$ width. Therefore an analysis of the $Z \gamma$ decay rate \cite{Gainer:2011aa} and its comparison with the $\gamma \gamma$ one, may reveal relevant properties of the possible new physics at the electroweak scale.  

Last but not the least, although the current data show a hint of an enhanced Higgs to diphoton decay width, we emphasize that  the analysis considered in this work could be applied  to the high energy and high luminosity data collected in the future.  If the Higgs boson is discovered, precision measurements of partial decay widths in various channels would become a top priority in order to properly identity the nature of the Higgs boson. In this case, the  current work will provide a guidance to place constraints on properties of possible new charged particles with significant couplings to the Higgs boson, which can contribute to the $\gamma\gamma$ and $Z\gamma$ widths.

 \begin {acknowledgements}
We would like to thank Bogdan Dobrescu, Stefania Gori, Ben Grinstein, Christophe Grojean, Nuria Rius and Nausheen Shah for discussions. Fermilab is operated by Fermi Research Alliance, LLC under Contract No. DE-AC02-07CH11359 with the U.S. Department of Energy. Work at ANL is supported in part by the U.S. Department of Energy under Contract No. DE-AC02-06CH11357. Work at Northwestern is supported in part by the U.S. Department of Energy under Contract No. DE-FG02-91ER40684. \end{acknowledgements}

 \section*{Appendix: Definitions of Loop Functions}
 Loop functions used in this paper are defined as follows:
 \bea
 \label{eq:loop1}
A_1(x)&=& -x^2\left[2x^{-2}+3x^{-1}+3(2x^{-1}-1)f(x^{-1})\right]\ .\\
\label{eq:loop2}
A_{1/2}(x) &=& 2  \, x^2 \left[x^{-1}+ (x^{-1}-1)f(x^{-1})\right] \ ,\\
 \label{eq:loop3}
A_0(x) &=& -x^2 \left[x^{-1}-f(x^{-1})\right]  \ , \\
\label{eq:loop4}
 A_1(x,y)&=& 4 (3-\tan^2\theta_w) I_2(x,y)+ \left[ (1+2\tau^{-1}) \tan^2\theta_w - (5+2\tau^{-1})\right] I_1(x,y) \ , \\
\label{eq:loop5}
 A_{1/2}(x,y)& =& I_1(x,y)-I_2(x,y) \ , \\ 
\label{eq:loop6}
A_{0}(x,y)& =& I_1(x,y) \ , 
 \eea
 where
 \bea
 I_1(x,y) &=& \frac{x y}{2(x-y)} + \frac{x^2 y^2}{2(x-y)^2}[ f(x^{-1})-f(y^{-1})] + \frac{x^2 y}{(x-y)^2}[g(x^{-1})-g(y^{-1})] \ ,\\
 I_2(x,y) &=& - \frac{x y}{2(x-y)} [ f(x^{-1})-f(y^{-1})]  \ .
 \eea
 It is worth pointing out that, compared with the definition in Ref.~\cite{Gunion:1989we}, we have factored the $ZWW$ triple gauge boson coupling in Eq.~(\ref{eq:tgc}) out of the loop function in $A_1(x,y)$.  For a Higgs mass below the kinematic threshold of the loop particle, $m_h < 2 \; m_{\rm loop}$, we have
 \bea
 f(x) &=& \arcsin^2 \sqrt{x} \ , \\
 g(x) &=& \sqrt{x^{-1} -1} \arcsin \sqrt{x} \ .
 \eea

\end{document}